# The Opposite of Smoothing: A Language Model Approach to Ranking Query-Specific Document Clusters

**Oren Kurland**  KURLAND@IE.TECHNION.AC.IL
**Eyal Krikon**  KRIKON@TX.TECHNION.AC.IL
*Faculty of Industrial Engineering and Management*
*Technion — Israel Institute of Technology*

## Abstract

Exploiting information induced from (*query-specific*) clustering of top-retrieved documents has long been proposed as a means for improving precision at the very top ranks of the returned results. We present a novel language model approach to ranking query-specific clusters by the presumed percentage of relevant documents that they contain. While most previous cluster ranking approaches focus on the cluster as a whole, our model utilizes also information induced from documents associated with the cluster. Our model substantially outperforms previous approaches for identifying clusters containing a high relevant-document percentage. Furthermore, using the model to produce document ranking yields precision-at-top-ranks performance that is consistently better than that of the initial ranking upon which clustering is performed. The performance also favorably compares with that of a state-of-the-art pseudo-feedback-based retrieval method.

## 1. Introduction

Users of search engines want to see the results most pertaining to their queries at the highest ranks of the returned document lists. However, attaining high precision at top ranks is still a very difficult challenge for search engines that have to cope with various (types of) queries (Buckley, 2004; Harman & Buckley, 2004).

High precision at top ranks is also important for applications that rely on search as an intermediate step; for example, question answering systems (Voorhees, 2002; Collins-Thompson, Callan, Terra, & Clarke, 2004). These systems have to provide an answer to a user's query rather than return a list of documents. Often, question answering systems employ a search over the given document corpus using the question at hand as a query (Voorhees, 2002). Then, passages of the highest ranked documents are analyzed for extracting (compiling) an answer to the question. Hence, it is important that the documents contain question-pertaining information.

To cope with the fact that a search engine can often return at the highest ranks of the result list quite a few documents that are not relevant to the user's query, researchers have proposed, among others, *cluster-based* result interfaces (Hearst & Pedersen, 1996; Leuski, 2001). That is, the documents that are initially highest ranked are clustered into clusters of similar documents. Then, the user can potentially exploit the clustering information to more quickly locate relevant documents from the initial result list. An important question in devising cluster-based result interfaces is the order by which to present the clusters to the users (Leuski, 2001). This order should potentially reflect the presumed percentage of relevant documents in the clusters.

Clusters of top-retrieved documents (a.k.a. *query-specific clusters*) can also be utilized without the user (or an application that uses search as an intermediate step) being aware that clustering has been performed. Indeed, researchers have proposed using information induced from the clusters to automatically re-rank the initially retrieved list so as to improve precision at top ranks (Preece, 1973; Willett, 1985; Hearst & Pedersen, 1996; Liu & Croft, 2004; Kurland & Lee, 2006; Yang, Ji, Zhou, Nie, & Xiao, 2006; Liu & Croft, 2008). Much of the motivation for employing clustering of top-retrieved documents comes from van Rijsbergen's *cluster hypothesis* (van Rijsbergen, 1979), which states that "closely associated documents tend to be relevant to the same requests". Indeed, it was shown that applying various clustering techniques to the documents most highly ranked by some initial search produces some clusters that contain a very high percentage of relevant documents (Hearst & Pedersen, 1996; Tombros, Villa, & van Rijsbergen, 2002; Kurland, 2006; Liu & Croft, 2006a). Moreover, positioning these clusters' constituent documents at the very top ranks of the returned results yields precision-at-top-ranks performance that is substantially better than that of state-of-the-art document-based retrieval approaches (Hearst & Pedersen, 1996; Tombros et al., 2002; Kurland, 2006).

Thus, whether used for creating effective result interfaces or for automatic re-ranking of search results, whether utilized so as to help serve users of search engines or applications that rely on search, query-specific clustering (i.e., clustering of top-retrieved documents) can result in much merit. Yet, a long standing challenge — progress with which can yield substantial retrieval effectiveness improvements over state-of-the-art retrieval approaches as we show — is the ability to identify query-specific clusters that contain a high percentage of documents relevant to the query.

We present a novel language-model-based approach to ranking query-specific clusters by the presumed percentage of relevant documents that they contain. The key insight that guides the derivation of our cluster-ranking model is that documents that are strongly associated with a cluster can serve as proxies for ranking it. Since documents can be considered as more focused units than clusters, they can serve, for example, as mediators for estimating the cluster-query "match". Thus, while most previous approaches to ranking various types of clusters focus on the cluster as a whole unit (Jardine & van Rijsbergen, 1971; Croft, 1980; Voorhees, 1985; Willett, 1985; Kurland & Lee, 2004; Liu & Croft, 2004, 2006b), our model integrates whole-cluster-based information with that induced from documents associated with the cluster. Hence, we conceptually take the opposite approach to that of cluster-based *smoothing* of document language models that has recently been proposed for document ranking (Azzopardi, Girolami, & van Rijsbergen, 2004; Kurland & Lee, 2004; Liu & Croft, 2004; Tao, Wang, Mei, & Zhai, 2006; Wei & Croft, 2006); that is, using cluster-based information to enrich a document representation for the purpose of document ranking.

Our model integrates two types of information induced from clusters and their proxy (associated) documents. The first is the estimated similarity to the query. The second is the *centrality* of an element (document or cluster) with respect to its reference set (documents in the initially-retrieved list or clusters of these documents); centrality is defined in terms of textual similarity to other central elements in the reference set (Kurland & Lee, 2005). Using either, or both, types of information just described — induced from a cluster as a whole and/or from its proxy documents — yields several novel cluster ranking criteria that are integrated in our model. We study the relative contribution of each of these criteria to the overall effectiveness of our approach. Furthermore, we show that previously proposed cluster ranking methods, which were developed independently, can be derived or explained using our ranking framework.

|  | AP | TREC8 | WSJ | WT10G |
|---|---|---|---|---|
| LM | 45.7 | 50.0 | 53.6 | 33.9 |
| Relevance model | 50.3 | 54.4 | 58.8 | 35.7 |
| Optimal cluster | **79.6** | **83.6** | **81.5** | **65.9** |

Table 1: The resultant p@5 performance of finding the "optimal cluster" in comparison to that of the initial LM-based ranking upon which clustering is performed, and that of an optimized relevance model.

Empirical evaluation shows that our cluster ranking model consistently and substantially outperforms previously proposed methods in identifying clusters that contain a high percentage of relevant documents. Furthermore, positioning the constituent documents of the cluster most highly ranked by our model at the top of the list of results yields precision-at-top-ranks performance that is substantially better than that of the initial document ranking upon which clustering was performed. The resultant performance also favorably compares with that of a state-of-the-art pseudo-feedback-based document retrieval method; and, with that of approaches that utilize inter-document similarities (e.g., using clusters) to directly (re-)rank documents.

## 2. Motivation

We first start by demonstrating the performance merits of the ability to effectively rank query-specific clusters by the presumed percentage of relevant documents that they contain.

Suppose that we have an initial list of documents that were retrieved in response to a query by using a standard language model (LM) approach (Ponte & Croft, 1998; Lafferty & Zhai, 2001). Suppose also that some clustering algorithm is used to cluster the 50 highest ranked documents, and that the resultant clusters contain 5 documents each. (Specific details of the experimental setup are provided in Section 5.2.) We define the *optimal cluster* as the one that contains the highest percentage of relevant documents. If we position the constituent documents of this cluster at the first five ranks of the document list returned in response to the query, then the resultant precision at 5 (p@5) performance is the percentage of relevant documents in this cluster. We contrast the p@5 performance with that of the initial LM-based ranking. As an additional reference comparison we use an optimized *relevance model*, RM3, which is a state-of-the-art pseudo-feedback-based query expansion approach (Lavrenko & Croft, 2001; Abdul-Jaleel et al., 2004). The performance numbers for four TREC corpora are presented in Table 1. (Details regarding the corpora and queries used are provided in Section 5.2.)

The message rising from Table 1 is clear. If we were able to automatically identify the optimal cluster, then the resultant performance would have been much better than that of the initial LM-based ranking upon which clustering is performed. Furthermore, the performance is also substantially better than that of a state-of-the-art retrieval approach. Similar conclusions were echoed in previous work on using clusters of top-retrieved documents (Hearst & Pedersen, 1996; Tombros et al., 2002; Crestani & Wu, 2006; Kurland, 2006; Liu & Croft, 2006a; Kurland & Domshlak, 2008).

## 3. Ranking Framework

Throughout this section we assume that the following have been fixed: a query $q$, a corpus of documents $\mathcal{D}$, and an initial list of $N$ documents $\mathcal{D}_{\text{init}}^N \subset \mathcal{D}$ (henceforth $\mathcal{D}_{\text{init}}$) that are the highest ranked by some search performed in response to $q$. We assume that $\mathcal{D}_{\text{init}}$ is clustered into a set of document clusters $Cl(\mathcal{D}_{\text{init}}) = \{c_1, \ldots, c_M\}$ by some clustering algorithm[1]. Our goal is to rank the clusters in $Cl(\mathcal{D}_{\text{init}})$ by the presumed percentage of relevant documents that they contain. In what follows we use the term "cluster" to refer either to the set of documents it is composed of, or to a (language) model induced from it. We use $p_y(x)$ to denote the language-model-based similarity between $y$ (a document or a cluster) and $x$ (a query or a cluster); we describe our language-model induction method in Section 5.1.

### 3.1 Cluster Ranking

Similarly to the language model approach to ranking documents (Ponte & Croft, 1998; Croft & Lafferty, 2003), and in deference to the recent growing interest in automatically labeling document clusters and topic models (Geraci, Pellegrini, Maggini, & Sebastiani, 2006; Treeratpituk & Callan, 2006; Mei, Shen, & Zhai, 2007), we state the problem of ranking clusters as follows: estimate the probability $p(c|q)$ that cluster $c$ can be labeled (i.e., its content can be described) by the terms in $q$. We hypothesize that the higher this probability is, the higher the percentage of documents pertaining to $q$ that $c$ contains.

Since $q$ is fixed, we use the rank equivalence

$$p(c|q) \stackrel{rank}{=} p(q|c) \cdot p(c)$$

to rank the clusters in $Cl(\mathcal{D}_{\text{init}})$. Thus, $c$ is ranked by combining the probability $p(q|c)$ that $q$ is "generated"[2] as a label for $c$ with $c$'s prior probability ($p(c)$) of "generating" *any* label. Indeed, most prior work on ranking various types of clusters (Jardine & van Rijsbergen, 1971; Croft, 1980; Willett, 1985; Voorhees, 1985; Kurland & Lee, 2004; Liu & Croft, 2004) implicitly uses uniform distribution for $p(c)$, and estimates $p(q|c)$ (in spirit) by comparing a representation of $c$ as a whole unit with that of $q$.

Here, we suggest to incorporate a document mediated approach to estimating the probability $p(q|c)$ of generating the label $q$ for cluster $c$. Since documents can be considered as more coherent units than clusters, they might help to generate more informative/focused labels than those generated by using representations of clusters as whole units. Such an approach is conceptually the opposite of *smoothing* a document representation (e.g., language model) with that of a cluster (Azzopardi et al., 2004; Kurland & Lee, 2004; Liu & Croft, 2004; Wei & Croft, 2006). In what follows we use $p(q|d)$ to denote the probability that $q$ is generated as a label describing document $d$'s content — cf., the language modeling approach to ranking documents (Ponte & Croft, 1998; Croft & Lafferty, 2003). Also, we assume that $p(d)$ — the prior probability that document $d$ generates any label — is a probability distribution over the documents in the corpus $\mathcal{D}$.

---

1. Clustering the documents most highly ranked by a search performed in response to a query is often termed *query-specific clustering* (Willett, 1985). We do not assume, however, that the clustering algorithm has knowledge of the query in hand.
2. While the term "generate" is convenient, we do not assume that clusters or documents literally generate labels, nor do we assume an underlying generative theory as that presented by Lavrenko and Croft (2001) and Lavrenko (2004), *inter alia*.

We let all, and only, documents in the corpus $\mathcal{D}$ to serve as proxies for label generation for any cluster in $Cl(\mathcal{D}_{\text{init}})$. Consequently, we assume that $p(d|c)$, the probability that $d$ is chosen as a proxy of $c$ for label generation, is a probability distribution defined over the documents in $\mathcal{D}$. Then, we can write using some probability algebra

$$p(c|q) \stackrel{rank}{=} p(c) \sum_{d_i \in \mathcal{D}} p(q|c, d_i) p(d_i|c). \tag{1}$$

We use $\lambda p(q|c) + (1-\lambda)p(q|d_i)$, where $\lambda$ is a free parameter, as an estimate for $p(q|c, d_i)$ (Si, Jin, Callan, & Ogilvie, 2002; Kurland & Lee, 2004) in Equation 1, and by applying probability algebra we get the following scoring principle[3] for clusters

$$\lambda p(c) p(q|c) + (1-\lambda) \sum_{d_i \in \mathcal{D}} p(q|d_i) p(c|d_i) p(d_i). \tag{2}$$

Equation 2 scores $c$ by a mixture of (i) the probability that $q$ is directly generated from $c$ combined with $c$'s prior probability of generating any label, and (ii) the (average) probability that $q$ is generated by documents that are both "strongly associated" with $c$ (as measured by $p(c|d_i)$) and that have a high prior probability $p(d_i)$ of generating labels.

We next derive specific ranking algorithms from Equation 2 by making some assumptions and estimation choices.

### 3.2 Algorithms

We first make the assumption, which underlies (in spirit) most pseudo-feedback-based retrieval models (Buckley, Salton, Allan, & Singhal, 1994; Xu & Croft, 1996; Lavrenko & Croft, 2003), that the probability of generating $q$ directly from $d_i$ ($p(q|d_i)$) is quite small for documents $d_i$ that are not in the initially retrieved list $\mathcal{D}_{\text{init}}$; hence, these documents have relatively little effect on the summation in Equation 2. Furthermore, if the clusters in $Cl(\mathcal{D}_{\text{init}})$ are produced by a "reasonable" clustering algorithm, then $p(c|d_i)$ — the cluster-document association strength — might be assumed to be significantly higher for documents from $\mathcal{D}_{\text{init}}$ that are in $c$ than for documents from $\mathcal{D}_{\text{init}}$ that are not in $c$. Consequently, we truncate the summation in Equation 2 by allowing only $c$'s constituent documents to serve as it proxies for generating $q$. Such truncation does not only alleviate the computational cost of estimating Equation 2, but can also yield improved effectiveness as we show in Section 5.3. In addition, we follow common practice in the language model framework (Croft

---

3. The shift in notation and terminology from "$p(c|q) \stackrel{rank}{=}$" to "score of $c$" echoes the transition from using (model) probabilities to estimates of such probabilities.

Kurland & Krikon

& Lafferty, 2003), specifically, in work on utilizing cluster-based language models for document retrieval (Liu & Croft, 2004; Kurland & Lee, 2004), and use language-model estimates for conditional probabilities to produce our primary ranking principle:

$$Score(c) \stackrel{def}{=} \lambda p(c)p_c(q) + (1-\lambda) \sum_{d_i \in c} p_{d_i}(q) p_{d_i}(c) p(d_i). \tag{3}$$

Note that using $p_d(c)$ for $p(c|d)$ means that we use the probability of generating the "label" $c$ (i.e., *some* term-based representation of $c$) from document $d$ as a surrogate for the document-cluster association strength.

The remaining task is to estimate the document and cluster priors, $p(d)$ and $p(c)$, respectively.

### 3.2.1 DOCUMENT AND CLUSTER BIASES

Following common practice in work on language-model-based retrieval we can use a uniform distribution for the document prior $p(d)$ (Croft & Lafferty, 2003), and similarly assume a uniform distribution for the cluster prior $p(c)$. Such practice would have been a natural choice if the clusters we want to rank were produced in a *query-independent* fashion. However, we would like to exploit the fact that the clusters in $Cl(\mathcal{D}_{\text{init}})$ are composed of documents in the initially retrieved list $\mathcal{D}_{\text{init}}$. A case in point, since $\mathcal{D}_{\text{init}}$ was retrieved in response to $q$, documents in $\mathcal{D}_{\text{init}}$ that are considered as reflecting $\mathcal{D}_{\text{init}}$'s content might be good candidates for generating the label $q$ (Kurland & Lee, 2005); a similar argument can be made for clusters in $Cl(\mathcal{D}_{\text{init}})$ that reflect its content. Therefore, instead of using "true" *prior* distributions, we use *biases* that represent the *centrality* (Kurland & Lee, 2005) of documents with respect to $\mathcal{D}_{\text{init}}$ and the centrality of clusters with respect to $Cl(\mathcal{D}_{\text{init}})$.[4]

We adopt a recently proposed approach to inducing document centrality that is based on measuring the similarity of a document in $\mathcal{D}_{\text{init}}$ to other central documents in $\mathcal{D}_{\text{init}}$ (Kurland & Lee, 2005). To quantify this recursive centrality definition, we compute PageRank's (Brin & Page, 1998) stationary distribution over a graph wherein vertices represent documents in $\mathcal{D}_{\text{init}}$ and edge-weights represent inter-document language-model-based similarities (Kurland & Lee, 2005). We then set $p(d) \stackrel{def}{=} Cent(d)$ for $d \in \mathcal{D}_{\text{init}}$ and 0 otherwise, where $Cent(d)$ is $d$'s PageRank score; hence, $p(d)$ is a probability distribution over the entire corpus $\mathcal{D}$.

Analogously, we set $p(c) \stackrel{def}{=} Cent(c)$ for $c \in Cl(\mathcal{D}_{\text{init}})$, where $Cent(c)$ is $c$'s PageRank score as computed over a graph wherein vertices are clusters in $Cl(\mathcal{D}_{\text{init}})$ and edge-weights represent language-model-based inter-cluster similarities; therefore, $p(c)$ is a probability distribution over the given set of clusters $Cl(\mathcal{D}_{\text{init}})$. The construction method of the document and cluster graphs follows that of constructing document-solely graphs (Kurland & Lee, 2005), and is elaborated in Appendix A.

Using the document and cluster induced biases we can now fully instantiate Equation 3 to derive **ClustRanker**, our primary cluster ranking algorithm:

---

4. The biases are not "true" *prior* distributions, because of the virtue by which $\mathcal{D}_{\text{init}}$ was created, that is, in response to the query. However, we take care that the biases form valid probability distributions as we show later.

| Algorithm | Scoring function ($Score(c)$) |
|---|---|
| ClustCent | $Cent(c)$ |
| ClustQueryGen | $p_c(q)$ |
| ClustCent $\wedge$ ClustQueryGen | $Cent(c)p_c(q)$ |
| DocCent | $\sum_{d_i \in c} p_{d_i}(c)Cent(d_i)$ |
| DocQueryGen | $\sum_{d_i \in c} p_{d_i}(q)p_{d_i}(c)$ |
| DocCent $\wedge$ DocQueryGen | $\sum_{d_i \in c} p_{d_i}(q)p_{d_i}(c)Cent(d_i)$ |
| ClustCent $\wedge$ DocCent | $\lambda Cent(c) + (1-\lambda)\sum_{d_i \in c} p_{d_i}(c)Cent(d_i)$ |
| ClustQueryGen $\wedge$ DocQueryGen | $\lambda p_c(q) + (1-\lambda)\sum_{d_i \in c} p_{d_i}(q)p_{d_i}(c)$ |
| ClustRanker | $\lambda Cent(c)p_c(q) + (1-\lambda)\sum_{d_i \in c} p_{d_i}(q)p_{d_i}(c)Cent(d_i)$ |

Table 2: Summary of methods for ranking clusters.

$$Score_{ClustRanker}(c) \stackrel{def}{=} \lambda Cent(c)p_c(q) + (1-\lambda)\sum_{d_i \in c} p_{d_i}(q)p_{d_i}(c)Cent(d_i). \qquad (4)$$

### 3.2.2 METHODS FOR RANKING CLUSTERS

The ClustRanker algorithm ranks cluster $c$ by integrating several criteria: (i) **ClustCent** — $c$'s centrality ($Cent(c)$), (ii) **ClustQueryGen** — the possibility to generate the label $q$ directly from $c$ as measured by $p_c(q)$, (iii) **DocCent** — the centrality of $c$'s constituent documents ($Cent(d)$), and (iv) **DocQueryGen** — the possibility to generate $q$ by $c$'s constituent documents as measured by $p_d(q)$. (Note that the latter two are combined with the cluster-document association strength, $p_d(c)$).

To study the effectiveness of each of these criteria (and some of their combinations) for ranking clusters, we apply the following manipulations to the ClustRanker algorithm: (i) setting $\lambda$ to 1 (0) to have only the cluster (documents) generate $q$, (ii) using uniform distribution for $Cent(c)$ (over $Cl(\mathcal{D}_{\text{init}})$) and/or for $Cent(d)$ (over $\mathcal{D}_{\text{init}}$) hence assuming that all clusters in $Cl(\mathcal{D}_{\text{init}})$ and/or documents in $\mathcal{D}_{\text{init}}$ are central to the same extent; we assume that the number of clusters in $Cl(\mathcal{D}_{\text{init}})$ is the same as the number of documents in $\mathcal{D}_{\text{init}}$, as is the case for the clustering method that we employ in Section 5; hence, the document uniform prior and the cluster uniform prior are the same; and (iv) setting $p_c(q)$ ($p_d(q)$) to the same constant value thereby assuming that for all clusters in $Cl(\mathcal{D}_{\text{init}})$ (documents in $\mathcal{D}_{\text{init}}$) the probability of directly generating $q$ is the same. For instance, setting $\lambda$ to 0 and $p_d(q)$ to some constant, we rank $c$ by DocCent — the weighted-average of the centrality values of its constituent documents: $\sum_{d_i \in c} p_{d_i}(c)Cent(d_i)$. Table 2 presents the resultant cluster ranking methods that we explore. ("$\wedge$" indicates that a method utilizes two criteria.)

### 3.3 Explaining Previous Methods for Ranking Clusters

The ClustRanker method, or more generally, Equation 3 on which it is based, can be used so as to help explain, and derive, some previously proposed methods for ranking clusters. While these methods were developed independently, and not as part of a single framework, their foundations can be described in terms of our approach. In the following discussion we use uniform prior for documents and clusters, and rank cluster $c$, using Equation 3, by $\lambda p_c(q) + (1-\lambda) \sum_{d_i \in c} p_{d_i}(q) p_{d_i}(c)$. Furthermore, recall that our framework is not committed to language models; i.e., $p_x(y)$, which is a language-model-based estimate for $p(y|x)$, can be replaced with another estimate.

Now, setting $\lambda = 1$, and consequently considering the cluster only as a whole unit, yields the most common cluster ranking method. That is, ranking the cluster based on the match of its representation as a whole unit with that of the query; the cluster can be represented, for example, by using the concatenation of its constituent documents (Kurland & Lee, 2004; Liu & Croft, 2004) or by a centroid-based representation of its constituent document representations (Voorhees, 1985; Leuski, 2001; Liu & Croft, 2008). The ClustQueryGen method from Table 2, which was also used in previous work (Liu & Croft, 2004; Kurland & Lee, 2004; Liu & Croft, 2006b; Kurland & Lee, 2006), is an example of this approach in the language modeling framework.

On the other hand, setting $\lambda = 0$ results in ranking $c$ by using its constituent documents rather than using $c$ as a whole unit: $\sum_{d_i \in c} p_{d_i}(q) p_{d_i}(c)$. Several cluster ranking methods that were proposed in past literature ignore the document-cluster association strength. This practice amounts to setting $p_{d_i}(c)$ to the same constant for all clusters and documents. Assuming also that all clusters contain the same number of documents, as is the case in our experimental setup in Section 5, we then rank $c$ by the arithmetic mean of the "query-match" values of its constituent documents, $\frac{1}{|c|} \sum_{d_i \in c} p_{d_i}(q)$; $|c|$ is the number of documents in $c$. The arithmetic mean can be bounded from above by $\max_{d_i \in c} p_{d_i}(q)$, which was used in some previous work on ranking clusters (Leuski, 2001; Shanahan, Bennett, Evans, Hull, & Montgomery, 2003; Liu & Croft, 2008), or from below by the geometric mean of the document-query match values, $\sqrt[|c|]{\prod_{d_i \in c} p_{d_i}(q)}$ (Liu & Croft, 2008; Seo & Croft, 2010).[5] Alternatively, the minimal query-document match value, $\min_{d_i \in c} p_{d_i}(q)$, which was also utilized for ranking clusters (Leuski, 2001; Liu & Croft, 2008), also constitutes a lower bound for the arithmetic mean.

## 4. Related Work

Query-specific clusters are often used to visualize the results of search so as to help users to quickly detect the relevant documents (Hearst & Pedersen, 1996; Leuski & Allan, 1998; Leuski, 2001; Palmer et al., 2001; Shanahan et al., 2003). Leuski (2001), for example, orders (hard) clusters in an interactive retrieval system by the highest and lowest query-similarity exhibited by any of their constituent documents. We showed in Section 3.3 that these ranking methods, and others, that were used in several reports on using query-specific clustering (Shanahan et al., 2003; Liu & Croft, 2006a), can be explained using our framework. Furthermore, in Section 5.3 we demonstrate the merits of ClustRanker with respect to these approaches.

Some work uses information from query-specific clusters to *smooth* language models of documents in the initial list so as to improve the document-query similarity estimate (Liu & Croft, 2004; Kurland, 2009). In a related vein, graph-based approaches for re-ranking the initial list, some using document clusters, that utilize inter-document similarity information were also proposed (Diaz, 2005; Kurland & Lee, 2005, 2006; Yang et al., 2006). These approaches can potentially help to improve the performance of our ClustRanker algorithm, as they provide a higher quality document ranking to begin with. Graph-based approaches for modeling inter-item textual similarities, some similar in spirit to our methods of inducing document and cluster centrality, were also used for text summarization, question answering, and clustering (Erkan & Radev, 2004; Mihalcea, 2004; Mihalcea & Tarau, 2004; Otterbacher, Erkan, & Radev, 2005; Erkan, 2006a, 2006b).

---

5. Liu and Croft (2008) and Seo and Croft (2010) used the geometric-mean-based language model representation of clusters, rather than the geometric mean of the "query-match" values.

Ranking query-specific (and query-independent) clusters in response to a query has traditionally been based on comparing a cluster representation with that of the query (Jardine & van Rijsbergen, 1971; Croft, 1980; Voorhees, 1985; Willett, 1985; Kurland & Lee, 2004; Liu & Croft, 2004, 2006b, 2006a). The ClustQueryGen criterion, which was used in work on ranking query-specific clusters in the language model framework (Liu & Croft, 2004; Kurland, 2009), is a language-model manifestation of this ranking approach. We show that the effectiveness of ClustQueryGen is inferior to that of ClustRanker in Section 5.3.

Some previous cluster-based document-ranking models (Kurland & Lee, 2004; Kurland, 2009) can be viewed as the conceptual opposite of our ClustRanker method as they use clusters as proxies for ranking documents. However, these models use only query-similarity information while ClustRanker integrates such information with centrality information. In fact, we show in Section 5.3 that centrality information is often more effective than query-similarity (generation) information for ranking query-specific clusters; and, that their integration yields better performance than that of using each alone.

Recently, researchers have identified some properties of query-specific clusters that contain a high percentage of relevant documents (Liu & Croft, 2006b; Kurland & Domshlak, 2008); among which are the cluster-query similarity (ClustQueryGen) (Liu & Croft, 2006b), the query similarity of the cluster's constituent documents (DocQueryGen) (Liu & Croft, 2006b; Kurland & Domshlak, 2008), and the differences between the two (Liu & Croft, 2006b). These properties were utilized for automatically deciding whether to employ cluster-based or document-based retrieval in response to a query (Liu & Croft, 2006b), and for ranking query-specific clusters (Kurland & Domshlak, 2008). The latter approach (Kurland & Domshlak, 2008) relies on rankings induced by clusters' models over the entire corpus, in contrast to our approach that focuses on the context within the initially retrieved list. However, our centrality-based methods from Table 2 can potentially be incorporated in this cluster-ranking framework (Kurland & Domshlak, 2008).

Some work on ranking query-specific clusters resembles ours in that it utilizes cluster-centrality information (Kurland & Lee, 2006); in contrast to our approach, centrality is induced based on cluster-document similarities. We further discuss this approach and compare it to ours in Section 5.3.

## 5. Evaluation

We next evaluate the effectiveness of our cluster ranking approach in detecting query-specific clusters that contain a high percentage of relevant documents.

### 5.1 Language-Model Induction

For language model induction, we treat documents and queries as term sequences. While there are several possible approaches of representing clusters as whole units (Voorhees, 1985; Leuski, 2001; Liu & Croft, 2006b; Kurland & Domshlak, 2008), our focus here is on the underlying principles of our ranking framework. Therefore, we adopt an approach commonly used in work on cluster-based retrieval (Kurland & Lee, 2004; Liu & Croft, 2004; Kurland & Lee, 2006; Liu & Croft, 2006a), and represent a cluster by the term sequence that results from concatenating its constituent documents. The order of concatenation has no effect since we only define unigram language models that assume term independence.

We use $p_x^{Dir[\mu]}(\cdot)$ to denote the Dirichlet-smoothed unigram language model induced from term sequence $x$ (Zhai & Lafferty, 2001); $\mu$ is the smoothing parameter. To avoid length bias and underflow issues when assigning language-model probabilities to long texts (Lavrenko et al., 2002; Kurland & Lee, 2005), as is the case for $p_d(c)$, we adopt the following measure (Lafferty & Zhai, 2001; Kurland & Lee, 2004, 2005, 2006), which is used for all the language-model-based estimates in the experiments to follow, unless otherwise specified (specifically, for relevance-model construction):

$$p_y(x) \stackrel{def}{=} \exp\left(-D\left(p_x^{Dir[0]}(\cdot) \,\middle\|\, p_y^{Dir[\mu]}(\cdot)\right)\right);$$

$x$ and $y$ are term sequences, and $D$ is the Kullback-Leibler (KL) divergence. The estimate was empirically demonstrated as effective in settings wherein long texts are assigned with language-model probabilities (Kurland & Lee, 2004, 2005, 2006).

Although the estimate just described does not constitute a probability distribution — as is the case for unigram language models — some previous work demonstrates the merits of using it as is without normalization (Kurland & Lee, 2005, 2006).

### 5.2 Experimental Setup

We conducted experiments with the following TREC corpora:

| corpus | # of docs | queries | disk(s) |
|--------|-----------|---------|---------|
| AP     | 242,918   | 51-64, 66-150 | 1-3 |
| TREC8  | 528,155   | 401-450 | 4-5 |
| WSJ    | 173,252   | 151-200 | 1-2 |
| WT10G  | 1,692,096 | 451-550 | WT10G |

Some of these data sets were used in previous work on ranking query-specific clusters (Liu & Croft, 2004; Kurland & Lee, 2006; Liu & Croft, 2008) with which we compare our methods. We used the titles of TREC topics for queries. We applied tokenization and Porter stemming via the Lemur toolkit (www.lemurproject.org), which was also used for language model induction.

We set $\mathcal{D}_{\text{init}}$, the list upon which clustering is performed, to the 50 highest ranked documents by an *initial ranking* induced over the entire corpus using $p_d^{Dir[\mu]}(q)$ — i.e., a standard language-model approach. To have an initial ranking of a reasonable quality, we set the smoothing parameter, $\mu$, to a value that results in optimized MAP (calculated at the standard 1000 cutoff) performance. This practice also facilitates the comparison with some previous work on cluster ranking (Kurland & Lee, 2006), which employs the same approach for creating an initial list of 50 documents to be clustered. The motivation for using a relatively short initial list rises from previous observations regarding the effectiveness of methods that utilize inter-document similarities among top-retrieved documents (Liu & Croft, 2004; Diaz, 2005; Kurland, 2006, 2009). The documents most highly ranked exhibit high query similarity, and hence, short retrieved lists could be viewed as providing a more "concise" corpus context for the query than longer lists. Similar considerations were echoed in work on pseudo-feedback-based query expansion, wherein top-retrieved documents are used for forming a new query model (Xu & Croft, 1996; Zhai & Lafferty, 2001; Lavrenko & Croft, 2001; Tao & Zhai, 2006).

To produce the set $Cl(\mathcal{D}_{\mathrm{init}})$ of query-specific clusters, we use a simple nearest-neighbors-based clustering approach that is known to produce (some) clusters that contain a high percentage of relevant documents (Kurland, 2006; Liu & Croft, 2006a). Given $d \in \mathcal{D}_{\mathrm{init}}$ we define a cluster that contains $d$ and the $k-1$ documents $d_i \in \mathcal{D}_{\mathrm{init}}$ ($d_i \neq d$) that yield the highest language-model similarity $p_{d_i}(d)$. (We break ties by document IDs.) The high percentages of relevant documents in an optimal cluster that were presented in Table 1 are for these clusters. More generally, this clustering approach was shown to be effective for cluster-based retrieval (Griffiths, Luckhurst, & Willett, 1986; Kurland & Lee, 2004; Kurland, 2006; Liu & Croft, 2006b, 2006a; Tao et al., 2006), specifically, with respect to using hard clusters (Kurland, 2009).

We posed our cluster ranking methods as a means for increasing precision at the very top ranks of the returned document list. Thus, we evaluate a cluster ranking method by the percentage of relevant documents in the highest ranked cluster. We use **p@k** to denote the percentage of relevant documents in a cluster of size $k$ (either 5 or 10), because it is the precision of the top $k$ documents that is obtained if the cluster's ($k$) constituent documents are positioned at the top ranks of the results. This cluster ranking evaluation approach was also employed in previous work on ranking clusters (Kurland & Lee, 2006; Liu & Croft, 2008) with which we compare our methods. We determine statistically significant differences of p@k performance using Wilcoxon's two-sided test at a confidence level of 95%.

To focus on the underlying principles of our approach and its potential effectiveness, and more specifically, to compare the relative effectiveness and contribution to the overall performance of the different information types utilized by our methods, we first ameliorate free-parameter-values effects. To that end, we set the values of free parameters incorporated by our methods to optimize *average* (over all queries per corpus) p@k performance for clusters of size $k$. (Optimization is based on a line search of the free-parameter values ranges.) We employ the same practice for *all* reference comparisons. That is, we independently optimize performance with respect to free-parameter values for p@5 and p@10. Then, in Section 5.3.5 we analyze the effect of free-parameter values on the effectiveness of our approach. In addition, in Section 5.3.6 we study the performance of our approach when free-parameter values are set using cross validation performed over queries. The value of $\lambda$, the interpolation parameter in the ClustRanker algorithm, is selected from $\{0, 0.1, \ldots, 1\}$. The values of the (two) parameters controlling the graph-construction methods (for inducing the document and cluster biases) are chosen from previously suggested ranges (Kurland & Lee, 2005). (See Appendix A for further details on graph construction.) The value of $\mu$, the language model smoothing parameter, is set to 2000 following previous recommenda-tions (Zhai & Lafferty, 2001), except for estimating $p_d(q)$ where we use the value chosen for creating $\mathcal{D}_{\text{init}}$ so as to maintain consistency with the initial ranking.

It is important to point out that the computational overhead of our approach on top of the initial search is not significant. Clustering of top-retrieved documents (50 in our case) can be performed quickly (Zamir & Etzioni, 1998); we note that our framework is not committed to a specific clustering approach. Furthermore, computing PageRank scores over a graph of 50 documents (clusters) to induce document (cluster) centrality takes only a few iterations of the Power method (Golub & Van Loan, 1996). Finally, we note that the number of documents in the corpus has no effect on the efficiency of our approach, as our methods are based on clustering the documents most highly ranked by the initial search.

### 5.3 Experimental Results

In what follows we present and analyze the performance numbers of our cluster ranking approach, and study the impact of various factors on its effectiveness. In Section 5.3.1 we study the effectiveness of ClustRanker as a means for improving precision at top ranks. To that end, we use a comparison with the initial ranking upon which clustering is performed, and with relevance models (Lavrenko & Croft, 2001). Then, in Section 5.3.2 we study the relative performance effect of the various cluster ranking criteria integrated by ClustRanker. We compare the effectiveness of ClustRanker with that of previously proposed methods for cluster ranking in Section 5.3.3. In Section 5.3.4 we compare the performance of ClustRanker with that of document-based re-ranking approaches that utilize inter-documents similarities in various ways. In Section 5.3.5 we analyze the performance sensitivity of ClustRanker with respect to free-parameter values. Finally, in Section 5.3.6 we analyze the performance of ClustRanker, and contrast it with that of various reference comparisons, when free-parameter values are set using cross validation performed over queries.

#### 5.3.1 Comparison with Document-Based Retrieval

The first question we are interested in is the effectiveness (or lack thereof) of ClustRanker in improving precision at the very top ranks. Recall that we use ClustRanker to rank clusters of $k$ ($\in \{5, 10\}$) documents from $\mathcal{D}_{\text{init}}$ — the initially retrieved document list. As described above, we evaluate ClustRanker's effectiveness by the percentage of relevant documents in the cluster most highly ranked. This percentage is the p@k attained if the cluster's constituent documents are positioned at the highest ranks of the final result list. In Table 3 we compare the performance of ClustRanker with that of the initial ranking. Since the initial ranking was created using a standard language-model-based document retrieval performed over the corpus with $p_d(q)$ as a scoring function, and with the document language model smoothing parameter ($\mu$) optimized for MAP, we also consider *optimized baselines* as reference comparisons: ranking all documents in the corpus by $p_d(q)$ where $\mu$ is set to optimize (independently) p@5 and p@10.

As we can see in Table 3, ClustRanker posts performance that is substantially better than that of the initial ranking in all relevant comparisons (corpus × evaluation measure). For AP and WT10G the performance improvements are also statistically significant. Furthermore, ClustRanker almost always outperforms the optimized baselines, often to a substantial extent; in several cases, the improvements are also statistically significant.

|  | AP || TREC8 || WSJ || WT10G ||
|---|---|---|---|---|---|---|---|---|
|  | p@5 | p@10 | p@5 | p@10 | p@5 | p@10 | p@5 | p@10 |
| init. rank. | 45.7 | 43.2 | 50.0 | 45.6 | 53.6 | 48.4 | 33.9 | 28.0 |
| opt. base. | 46.5 | 43.7 | 51.2 | 46.4 | **56.0** | 49.4 | 34.1 | 28.2 |
| ClustRanker | **52.7**$^i$ | **50.6**$^i_o$ | **57.6** | **50.6** | **56.0** | **51.2** | **39.8**$^i_o$ | **33.9**$^i_o$ |

Table 3: Comparison of ClustRanker with the initial document ranking and optimized baselines. Boldface marks the best result in a column; 'i' and 'o' mark statistically significant differences with the initial ranking, and optimized baselines, respectively.

|  | AP || TREC8 || WSJ || WT10G ||
|---|---|---|---|---|---|---|---|---|
|  | p@5 | p@10 | p@5 | p@10 | p@5 | p@10 | p@5 | p@10 |
| init. rank. | 45.7 | 43.2 | 50.0 | 45.6 | 53.6 | 48.4 | 33.9 | 28.0 |
| Rel Model | 50.3$^i$ | 48.6$^i$ | 54.4 | 50.2 | **58.4**$^i$ | **53.2**$^i$ | 35.7 | 29.9 |
| Rel Model(Re-Rank) | 51.1$^i$ | 48.3$^i$ | 53.6 | 49.8 | 58.8$^i$ | 53.4$^i$ | 36.3 | 30.1 |
| ClustRanker | **52.7**$^i$ | **50.6**$^i$ | **57.6** | **50.6** | 56.0 | 51.2 | **39.8**$^i$ | **33.9**$^i$ |

Table 4: Comparison of ClustRanker with a relevance model (RM3) used to either rank the entire corpus (Rel Model) or to re-rank the initial list (Rel Model(Re-Rank)). Boldface marks the best result in a column; 'i' marks statistically significant difference with the initial ranking.

**Comparison with Pseudo-Feedback-Based Retrieval**  The ClustRanker algorithm helps to identify relevant documents in $\mathcal{D}_{\text{init}}$ by exploiting clustering information. Pseudo-feedback-based query expansion approaches, on the other hand, define a query model based on $\mathcal{D}_{\text{init}}$ and use it for (re-)ranking the entire corpus (Buckley et al., 1994; Xu & Croft, 1996). To contrast the two paradigms, we use the *relevance model* RM3 (Lavrenko & Croft, 2001; Abdul-Jaleel et al., 2004; Diaz & Metzler, 2006), which is a state-of-the-art pseudo-feedback-based query expansion approach. We use RM3 for ranking the entire corpus as is standard, and refer to this implementation as **Rel Model**. Since ClustRanker can be thought of as a means to re-ranking the initial list $\mathcal{D}_{\text{init}}$, we also experiment with using RM3 for re-ranking only $\mathcal{D}_{\text{init}}$, rather than the entire corpus; **Rel Model(Re-Rank)** denotes this implementation. We set the values of the free parameters of Rel Model and Rel Model(Re-Rank) so as to independently optimize p@5 and p@10 performance. (See Appendix B for details regarding the relevance model implementation.)

We can see in Table 4 that ClustRanker outperforms the relevance models on AP, TREC8 and WT10G; for WSJ, the relevance models outperform ClustRanker. The performance differences between ClustRanker and the relevance models, however, are not statistically significant. Nevertheless, these results attest to the overall effectiveness of our approach in attaining high precision at top ranks. As we later show, previous methods for ranking clusters often yield performance that is only comparable to that of the initial ranking, and much inferior to that of the relevance model.

|  | AP | | TREC8 | | WSJ | | WT10G | |
|---|---|---|---|---|---|---|---|---|
|  | p@5 | p@10 | p@5 | p@10 | p@5 | p@10 | p@5 | p@10 |
| init. rank. | 45.7 | 43.2 | 50.0 | 45.6 | 53.6 | 48.4 | 33.9 | 28.0 |
| ClustCent | 51.7 | 48.6$^i$ | 52.4 | 49.4 | 54.8 | 50.0 | **39.8**$^i$ | 33.0$^i$ |
| ClustQueryGen | 39.2$^i$ | 38.8$^i$ | 39.6$^i$ | 40.6$^i$ | 44.0$^i$ | 37.0$^i$ | 30.0 | 24.1 |
| ClustCent ∧ ClustQueryGen | 49.7 | 48.0$^i$ | 55.2 | 50.4 | 52.4 | 47.8 | 39.6$^i$ | 33.1$^i$ |
| DocCent | 52.9$^i$ | 48.8 | 52.0 | 48.8 | 55.6 | 50.6 | 31.0 | 28.1 |
| DocQueryGen | 43.6 | 46.7 | 47.6 | 43.2 | 55.2 | 47.0 | 33.5 | 27.0 |
| DocCent ∧ DocQueryGen | 52.7$^i$ | **50.6**$^i$ | 54.8 | 49.0 | **56.0** | 51.2 | 37.1 | 31.4 |
| ClustCent ∧ DocCent | **53.5**$^i$ | 48.8$^i$ | 54.8 | 49.8 | **56.0** | 51.4 | **39.8**$^i$ | 33.0$^i$ |
| ClustQueryGen ∧ DocQueryGen | 43.6 | 46.7 | 47.6 | 43.2 | 55.2 | 47.8 | 36.5 | 29.1 |
| ClustRanker | 52.7$^i$ | **50.6**$^i$ | **57.6** | **50.6** | **56.0** | 51.2 | **39.8**$^i$ | **33.9**$^i$ |

Table 5: Comparison of the cluster ranking methods from Table 2. Boldface marks the best result in a column and 'i' indicates a statistically significant difference with the initial ranking.

### 5.3.2 Deeper Inside ClustRanker

We now turn to analyze the performance of the various cluster ranking criteria (methods) that ClustRanker integrates so as to study their relative contribution to its overall effectiveness. (Refer back to Table 2 for specification of the different methods.) The performance numbers are presented in Table 5.

We are first interested in the comparison of the two types of information utilized for ranking, that is, centrality and query-similarity (generation). We can see in Table 5 that in almost all relevant comparisons (corpus × evaluation metric), using centrality information yields performance that is superior to that of using query-similarity (generation) information. (Compare ClustCent with ClustQueryGen, DocCent with DocQueryGen, and ClustCent ∧ DocCent with ClustQueryGen ∧ DocQueryGen.) Specifically, we see that cluster-query similarity (ClustQueryGen), which was the main ranking criterion in previous work on cluster ranking, yields performance that is much worse than that of cluster centrality (ClustCent) — a cluster ranking criterion which is novel to this study. In addition, we note that integrating centrality and query-similarity (generation) information can often yield performance that is better than that of using each alone, as is the case for DocCent ∧ DocQueryGen with respect to DocCent and DocQueryGen.

We next turn to examine the relative effectiveness of using the cluster as a whole versus using its constituent documents. When using only query-similarity (generation) information, we see that using the documents in the cluster is much more effective than using the cluster as a whole. (Compare DocQueryGen with ClustQueryGen.) This finding further attests to the merits of using documents as proxies for ranking clusters — the underlying idea of our approach. When using centrality information, the picture is split across corpora: for AP and WSJ using the documents in the cluster yields better performance than using the whole cluster, while the reverse holds for TREC8 and WT10G. (Compare DocCent with ClustCent.) Integrating whole-cluster-based and document-based information results in performance that is for all corpora (much) better than that of using the less effective of the two, and sometimes even better than the more effective of the two.

|  | AP | | TREC8 | | WSJ | | WT10G | |
|---|---|---|---|---|---|---|---|---|
|  | p@5 | p@10 | p@5 | p@10 | p@5 | p@10 | p@5 | p@10 |
| init. rank. | 45.7 | 43.2 | 50.0 | 45.6 | 53.6 | 48.4 | 33.9 | 28.0 |
| $d \in \mathcal{D}_{\text{init}}$ | 49.5 | 47.6 | 54.0 | 49.8 | 52.8 | 49.6 | $39.6^i$ | $33.2^i$ |
| $d \in c$ | $\mathbf{52.7^i}$ | $\mathbf{50.6^i}$ | **57.6** | **50.6** | **56.0** | **51.2** | $\mathbf{39.8^i}$ | $\mathbf{33.9^i}$ |

Table 6: Performance numbers of ClustRanker when either *all* documents in $\mathcal{D}_{\text{init}}$ serve as proxies for cluster $c$ (denoted $d \in \mathcal{D}_{\text{init}}$), or only $c$'s constituent documents serve as its proxies, as in the original implementation (denoted $d \in c$). Boldface marks the best result in a column; 'i' marks statistically significant differences with the initial ranking.

It is not a surprise, then, that the ClustRanker method, which integrates centrality information and query-similarity (generation) information that are induced from both the cluster as a whole and from its constituent documents, is in most relevant comparisons the most effective cluster ranking method among those presented in Table 5.

**Documents as Proxies for Clusters** The findings presented above demonstrated the merits of using documents as proxies for clusters. We now turn to study the effect on performance of the documents selected as proxies. The derivation of ClustRanker was based on truncating the summation in Equation 2 (Section 3) so as to allow only $c$'s constituent documents to serve as its proxies. We examine a variant of ClustRanker wherein *all* documents in the initial list $\mathcal{D}_{\text{init}}$ can serve as $c$'s proxies:

$$Score(c) \stackrel{def}{=} Cent(c)p_c(q) + (1-\lambda) \sum_{d_i \in \mathcal{D}_{\text{init}}} p_{d_i}(q) p_{d_i}(c) Cent(d_i).$$

As can be seen in Table 6, this variant (represented by the row labeled "$d \in \mathcal{D}_{\text{init}}$") posts performance that is almost always better than that of the initial document ranking from which $\mathcal{D}_{\text{init}}$ is derived. However, the performance is also consistently worse than that of the original implementation of ClustRanker (represented by the row labeled "$d \in c$") that lets only $c$'s constituent documents to serve as its proxies. Furthermore, this variant of ClustRanker posts less statistically significant improvements over the initial ranking than the original implementation. (The performance differences between the two variants of ClustRanker, however, are not statistically significant.) Thus, as was mentioned in Section 3, using only the cluster's constituent documents as its proxies is not only computationally convenient, but also yields performance improvements.

5.3.3 COMPARISON WITH PAST APPROACHES FOR RANKING CLUSTERS

In Table 7 we compare the performance of ClustRanker with that of previously proposed methods for ranking clusters. In what follows we first discuss these methods, and then analyze the performance patterns.

Most previous approaches to ranking (various types of) clusters compare a cluster representation with that of the query (Jardine & van Rijsbergen, 1971; Croft, 1980; Kurland & Lee, 2004; Liu & Croft, 2004, 2006b). Specifically, in the language model framework, query-specific clusters were ranked by the probability assigned by their induced language models to the query (Liu & Croft, 2004, 2006b; Kurland & Lee, 2006). Note that this is exactly the ClustQueryGen method in our setup, which ranks $c$ by $p_c(q)$.

There has been some work on using the maximal (Leuski, 2001; Shanahan et al., 2003; Liu & Croft, 2008) and minimal (Leuski, 2001; Liu & Croft, 2008) query-similarity values of the documents in a cluster for ranking it. We showed in Section 3.3 that these approaches can be explained in terms of our framework; specifically, the score of cluster $c$ is $\max_{d_i \in c} p_{d_i}(q)$ and $\min_{d_i \in c} p_{d_i}(q)$, respectively. However, these methods were originally proposed for ranking hard clusters. As the clusters we rank here are overlapping, these ranking criteria are somewhat less appropriate as they result in many ties of cluster scores. Still, we use these methods as baselines and break ties arbitrarily as they were also used in some recent work on ranking nearest-neighbors-based clusters (Liu & Croft, 2008).[6] Following some observations with regard to the merits of representing clusters by using the geometric mean of their constituent documents' representations (Liu & Croft, 2008; Seo & Croft, 2010), we also consider the geometric mean of the query-similarity values of documents in $c$ for ranking it; that is, $\sqrt[|c|]{\prod_{d_i \in c} p_{d_i}(q)}$.[7]

An additional reference comparison that we consider, which was shown to yield effective cluster ranking performance, is a recently proposed (bipartite-)graph-based approach (Kurland & Lee, 2006). Documents in $\mathcal{D}_{\text{init}}$ are vertices on one side, and clusters in $Cl(\mathcal{D}_{\text{init}})$ are vertices on the other side; an edge connects document $d$ with the $\delta$ clusters $c_i$ that yield the highest language-model similarity $p_{c_i}(d)$, which also serves as a weight function for the edges. Then, Kleinberg's (1997) **HITS** (hubs and authorities) algorithm is run on the graph, and clusters are ranked by their induced *authority* values. It was shown that the cluster with the highest authority value tends to contain a high percentage of relevant documents (Kurland & Lee, 2006). For implementation, we follow the details provided by Kurland and Lee (2006); specifically, we choose the value of $\delta$ from $\{2, 4, 9, 19, 29, 39, 49\}$ so as to optimize p@k performance for clusters of size $k$.

Table 7 presents the comparison of ClustRanker with the reference comparisons just described. The p@k ($k \in \{5, 10\}$) reported for a cluster ranking method is the percentage of relevant documents in the highest ranked cluster, wherein clusters contain $k$ documents each.

We can see in Table 7 that ClustRanker outperforms all reference comparisons in almost all cases. Many of the these performance differences are also statistically significant. Furthermore, ClustRanker is the only cluster ranking method in Table 7 that consistently outperforms the initial ranking. Moreover, ClustRanker posts more statistically significant improvements over the initial ranking than the other cluster ranking methods do.

---

6. As noted above, previous work on cluster-based retrieval has demonstrated the merits of using overlapping nearest-neighbors-based clusters with respect to using hard clusters (Kurland, 2009). Indeed, recent work on cluster ranking has focused on ranking nearest-neighbor-based clusters as we do here (Kurland & Lee, 2006; Liu & Croft, 2006b, 2006a, 2008; Seo & Croft, 2010).
7. We note that in the original proposals (Liu & Croft, 2008; Seo & Croft, 2010) the geometric mean of language models was used at the term level rather than at the query-assigned score level as we use here. To maintain consistency with the other cluster-ranking methods explored, we use the geometric mean at the query-assigned score level; and, we hasten to point out that a geometric-mean-based language model (Liu & Croft, 2008; Seo & Croft, 2010) could be used instead of the standard language model for clusters in ClustRanker and in the reference comparisons so as to potentially improve performance.

|  | AP | | TREC8 | | WSJ | | WT10G | |
|---|---|---|---|---|---|---|---|---|
|  | p@5 | p@10 | p@5 | p@10 | p@5 | p@10 | p@5 | p@10 |
| init. rank. | 45.7 | 43.2 | 50.0 | 45.6 | 53.6 | 48.4 | 33.9 | 28.0 |
| $S_{ClustQueryGen}(c) \stackrel{def}{=} p_c(q)$ | $39.2^i$ | $38.8^i$ | $39.6^i$ | $40.6^i$ | $44.0^i$ | $37.0^i$ | 30.0 | 24.1 |
| $S_{Max}(c) \stackrel{def}{=} \max_{d_i \in \mathcal{D}_{init}} p_{d_i}(q)$ | 41.8 | 40.3 | $38.8^i$ | 41.6 | 51.2 | 46.6 | 33.9 | 29.2 |
| $S_{Min}(c) \stackrel{def}{=} \min_{d_i \in \mathcal{D}_{init}} p_{d_i}(q)$ | 47.0 | 46.7 | 46.4 | 48.4 | 48.4 | 47.8 | 31.4 | 25.9 |
| $S_{GeoMean}(c) \stackrel{def}{=} \sqrt[|c|]{\prod_{d_i \in c} p_{d_i}(q)}$ | 44.4 | $46.7^i$ | 50.0 | 49.6 | **56.0** | 50.6 | 37.4 | $31.8^i$ |
| $S_{HITS}(c)$ | 49.5 | 47.2 | 50.8 | 46.6 | 53.6 | 49.0 | $26.7^i$ | $23.9^i$ |
| $S_{ClustRanker}(c)$ | $\mathbf{52.7}^{icM}_{g}$ | $\mathbf{50.6}^{icM}$ | $\mathbf{57.6}^{cM}_{mh}$ | $\mathbf{50.6}^{cM}_{m}$ | $\mathbf{56.0}^{c}_{m}$ | $\mathbf{51.2}^{c}$ | $\mathbf{39.8}^{icM}_{mh}$ | $\mathbf{33.9}^{icM}_{mh}$ |

Table 7: Comparison of ClustRanker with previously proposed methods for ranking clusters. ('$S$' stands for the cluster-scoring function.) Boldface marks the best result in a column; 'i' marks statistically significant difference between a method and the initial ranking; 'c', 'M', 'm', 'g', and 'h' mark statistically significant differences of ClustRanker with the ClustQueryGen, Max, Min, GeoMean and HITS methods, respectively.

|  | AP | | TREC8 | | WSJ | | WT10G | |
|---|---|---|---|---|---|---|---|---|
|  | p@5 | p@10 | p@5 | p@10 | p@5 | p@10 | p@5 | p@10 |
| init. rank. | 45.7 | 43.2 | 50.0 | 45.6 | 53.6 | 48.4 | 33.9 | 28.0 |
| HITS | 49.5 | 47.2 | 50.8 | 46.6 | 53.6 | 49.0 | $26.7^i$ | $23.9^i$ |
| ClustCent | 51.7 | $48.6^i$ | 52.4 | 49.4 | 54.8 | 50.0 | $\mathbf{39.8}^{ih}$ | $\mathbf{33.0}^{ih}$ |
| DocCent | $52.9^{ih}$ | **48.8** | 52.0 | 48.8 | 55.6 | 50.6 | $31.0^h$ | $28.1^h$ |
| ClustCent $\wedge$ DocCent | $\mathbf{53.5}^{ih}$ | $\mathbf{48.8}^i$ | **54.8** | **49.8** | **56.0** | **51.4** | $\mathbf{39.8}^{ih}$ | $\mathbf{33.0}^{ih}$ |

Table 8: Comparison of our centrality-solely approaches for ranking clusters with the HITS-based method (Kurland & Lee, 2006). Boldface marks the best performance in a column; 'i' and 'h' mark statistically significant differences with the initial ranking and the HITS method, respectively.

**The Comparison with the HITS-Based Approach**  The HITS-based method (Kurland & Lee, 2006) utilizes cluster centrality information as induced over a cluster-document graph. Our ClustRanker method, on the other hand, integrates centrality information — induced over document-solely and cluster-solely graphs — with query-similarity (generation) information. In Table 8 we contrast the resultant performance of using the different notions of centrality utilized by the two methods. We present the performance of our centrality-solely-based methods ClustCent, DocCent, and ClustCent $\wedge$ DocCent and of the HITS approach (Kurland & Lee, 2006).

We can see in Table 8 that all our centrality-solely-based approaches outperform the HITS-based method in all relevant comparisons. These results attest to the effective utilization of (a specific type of) centrality information by our framework.

### 5.3.4 Comparison with Utilizing Inter-Document Similarities to Directly Rank Documents

Ranking clusters based on the presumed percentage of relevant documents that they contain, as in ClustRanker, is one approach of utilizing inter-document similarities so as to improve document-ranking effectiveness. Alternatively, inter-document similarities can be exploited so as to directly rank documents. For example, a re-ranking principle that was demonstrated to be effective is rewarding documents that are initially highly ranked, and which are highly similar to many other documents in the initial list (Baliński & Daniłowicz, 2005; Diaz, 2005; Kurland & Lee, 2005). Specifically, using the DocCent component of ClustRanker, that is, the PageRank score of document $d$ ($Cent(d)$) as induced over a document-similarity graph, and scaling this value by $p_d(q)$ — which is the initial (query similarity) score of $d$ — was shown to be an effective re-ranking criterion (Kurland & Lee, 2005). We use **PR+QuerySim** to denote this method.

Another approach that we consider is ranking documents using clusters as a form of an "extended" document representation. (In language model terms this translates to cluster-based smoothing.) Specifically, we use the **Interpolation** algorithm (Kurland & Lee, 2004), which was shown to be highly effective for re-ranking (Kurland, 2009). A document $d$ is scored by $\lambda p_d(q) + (1 - \lambda) \sum_{c \in Cl(\mathcal{D}_{\text{init}})} p_c(q) p_d(c)$; $\lambda$ is a free parameter. That is, a document is rewarded by its "direct match" with the query ($p_d(q)$), which is the criterion used for creating the initial ranking, and by the "query match" of clusters ($p_c(q)$) with which it is strongly associated (as measured by $p_d(c)$). In other words, the Interpolation model backs off from a document representation to a cluster-based representation. The Interpolation model could conceptually be viewed as a generalization of methods that use a single cluster (Liu & Croft, 2004; Tao et al., 2006) or a topic model (Wei & Croft, 2006) for smoothing a document language model. Furthermore, note that while Interpolation uses clusters as document proxies for ranking documents, our ClustRanker method uses documents as cluster proxies for ranking clusters.

In Table 9 we compare the performance of ClustRanker with that of the PR+QuerySim and Interpolation methods just described. The performance of Interpolation is independently optimized for p@5 and p@10 using clusters of 5 and 10 documents, respectively, as is the case for ClustRanker; the value of $\lambda$ is chosen from $\{0, 0.1, \ldots, 0.9\}$. The performance of PR+QuerySim is also independently optimized for p@5 and p@10, when setting the graph out degree parameter ($\delta$) to 4 and 9 respectively, and selecting the value of $\nu$ from $\{2, 4, 9, 19, 29, 39, 49\}$. (Setting the graph out-degree parameter to a value $x$ amounts to having a document "transfer" centrality support to $x$ other documents; hence, using the values of 4 and 9 amounts to considering "local neighborhoods" of 5 and 10 documents in the similarity space, respectively; this is conceptually reminiscent of using clusters of size 5 and 10 respectively. Refer to Appendix A for further details.)

We can see in Table 9 that ClustRanker outperform PR+QuerySim and Interpolation in most relevant comparisons. (Several of the performance differences with the former are statistically significant, while those with the latter are not.) Specifically, the relative performance improvements for WT10G are quite substantial. (We hasten to point out, however, that Interpolation posts more statistically significant performance improvements over the initial ranking than ClustRanker does.)

|  | AP | | TREC8 | | WSJ | | WT10G | |
|---|---|---|---|---|---|---|---|---|
|  | p@5 | p@10 | p@5 | p@10 | p@5 | p@10 | p@5 | p@10 |
| init. rank. | 45.7 | 43.2 | 50.0 | 45.6 | 53.6 | 48.4 | 33.9 | 28.0 |
| PR+QuerySim | 49.5 | $49.5^i$ | 56.0 | $\mathbf{51.0}^i$ | $\mathbf{57.2}^i$ | 50.4 | 35.9 | 30.4 |
| Interpolation | $51.3^i$ | $50.3^i$ | $55.6^i$ | $49.6^i$ | 56.8 | $\mathbf{52.4}$ | 36.1 | $31.8^i$ |
| ClustRanker | $\mathbf{52.7}^i_p$ | $\mathbf{50.6}^i$ | $\mathbf{57.6}$ | 50.6 | 56.0 | 51.2 | $\mathbf{39.8}^i_p$ | $\mathbf{33.9}^i_p$ |

Table 9: Comparison with re-ranking methods that utilize inter-document similarities to directly rank documents. Boldface marks the best result in a column. Statistically significant differences with the initial ranking and PR+QuerySim are marked with 'i' and 'p', respectively. There are no statistically significant differences between ClustRanker and Interpolation.

All in all, we see that ranking clusters as is done by ClustRanker can result in (document) re-ranking performance that is at least as effective (and often more effective) than that of methods that utilize inter-document similarities to directly rank documents.

5.3.5 PERFORMANCE-SENSITIVITY ANALYSIS

We next turn to analyze the effect of varying the values of the free parameters that ClustRanker incorporates on its performance. The first parameter, $\lambda$, controls the reliance on the cluster as a whole unit versus its constituent documents. (Refer back to Equation 4 in Section 3 for details.) Figure 1 depicts the p@5 performance of ClustRanker, with clusters of 5 documents, as a function of $\lambda$; the p@10 performance patterns of ClustRanker with clusters of 10 documents are similar, and are omitted to avoid cluttering the presentation.

We can see in Figure 1 that for AP, TREC8 and WT10G, all values of $\lambda$ result in performance that is (much) better than that of the initial ranking; for WSJ, $\lambda \leq 0.4$ results in performance superior to that of the initial ranking. Furthermore, for AP, TREC8, and WT10G, $\lambda = 0.4$, which strikes a good balance between using the cluster as a whole and using its constituent documents, yields (near) optimal performance; for WSJ, smaller values of $\lambda$ (specifically, $\lambda = 0$), which result in more weight put on the cluster's constituent documents, are more effective. Thus, we witness again the importance of using the cluster's constituent documents as proxies when ranking the cluster.

The graph-based method used by ClustRanker for inducing document (and cluster) centrality depends on two free parameters: $\delta$ (the number of nearest neighbors considered for each element in the graph), and $\nu$ (PageRank's damping factor); see Appendix A for details. As noted above, both the document graph and the cluster graph are constructed using the same values of these two parameters. Figures 2 and 3 depict the effect of the values of $\delta$ and $\nu$, respectively, on the p@5 performance of ClustRanker with clusters of 5 documents.

We can see in Figure 2 that all values of $\delta$ result in performance that is (often much) better than that of the initial ranking. In general, small values of $\delta$ yield the best performance. This finding is in accordance with those reported in previous work on using nearest-neighbor-based graphs for ranking documents in an initially retrieved list using document-solely graphs (Diaz, 2005; Kurland & Lee, 2005).

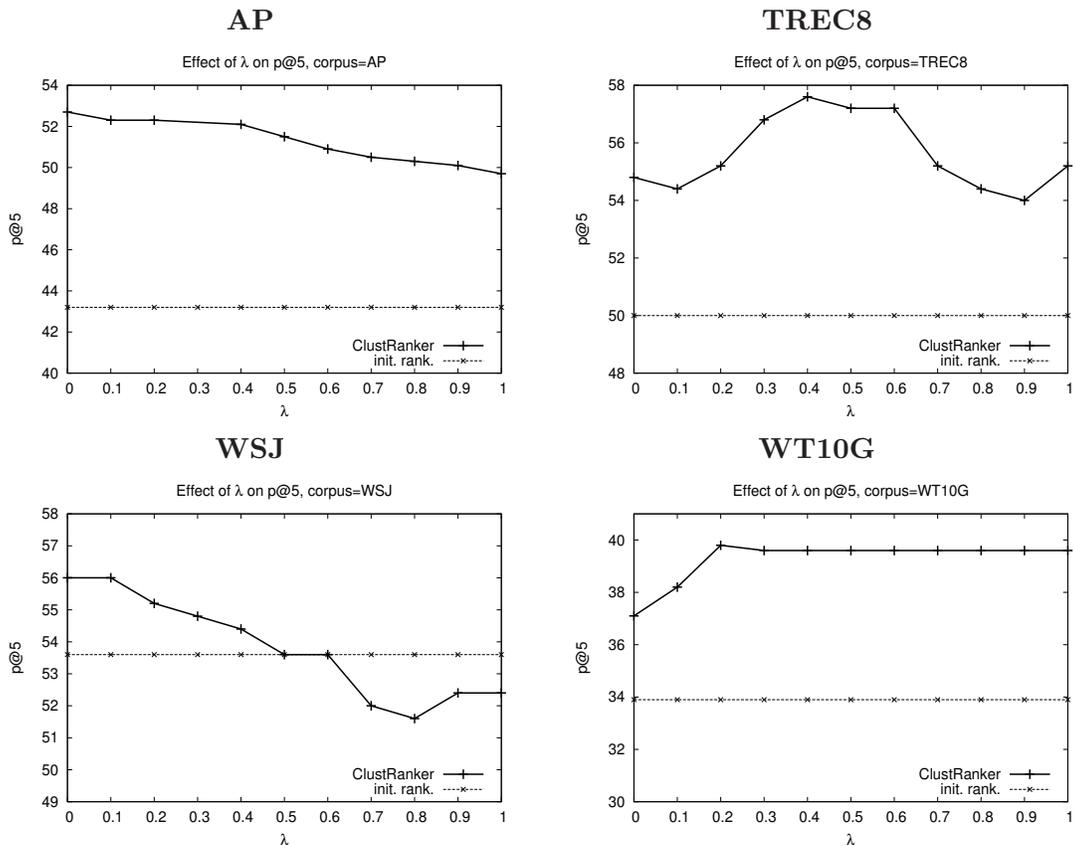

Figure 1: Effect of varying $\lambda$ on the p@5 performance of ClustRanker.

Figure 3 shows that for almost every value of $\nu$, the resultant performance of ClustRanker transcends that of the initial ranking; in many cases, the improvement is quite substantial.

### 5.3.6 Learning Free-Parameter Values

Heretofore, we studied the performance of ClustRanker, and analyzed the effectiveness of the information types that it utilizes, while ameliorating free-parameter values effects. That is, we reported performance for parameter values that result in optimized average p@k over the entire set of queries per corpus. We have applied the same practice for all reference comparisons that we considered, which resulted in comparing the potential effectiveness of our approach with that of previously suggested ones. In addition, we studied in the previous section the effect of the values of free parameters incorporated by ClustRanker on its (average) performance.

We now turn to study the question of whether effective values of the free parameters of ClustRanker generalize across queries; that is, whether these values can be learned. To perform this study, we employ a leave-one-out cross validation procedure. For each query, the free parameters of ClustRanker are set to values that yield optimal average p@k over all *other* queries for the same corpus. Then, we report the resultant average p@k over all queries per corpus. Thus, the reported p@k numbers are based on learning performed with p@k as the optimization metric.

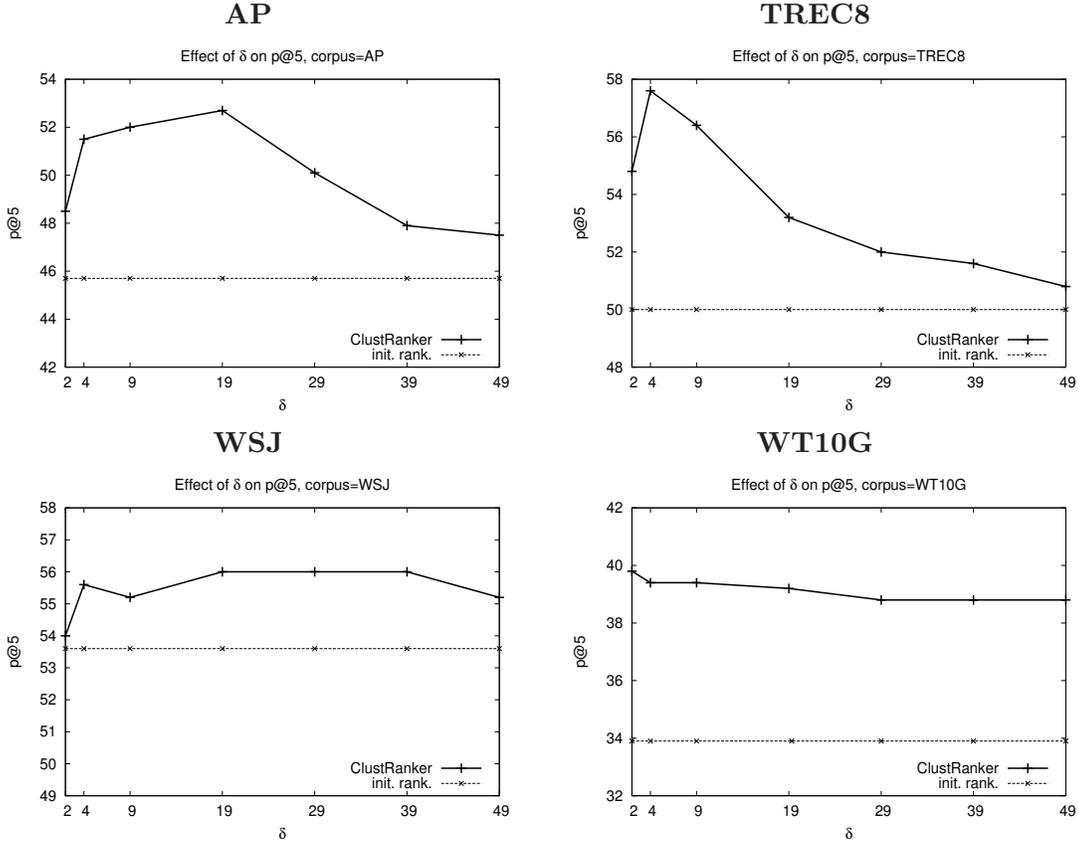

Figure 2: Effect of varying $\delta$, one of the graph parameters (refer to Appendix A), on the p@5 performance of ClustRanker.

We contrast the performance of ClustRanker with that of the reference comparisons used above. Specifically, the document-based ranking baselines: (i) the initial ranking, (ii) the relevance model used to rank the entire corpus (Rel Model), (iii) the relevance model used to re-rank the initial list (Rel Model(Re-Rank)); and, the cluster ranking methods: (i) $Score_{ClustQueryGen}(c) \stackrel{def}{=} p_c(q)$, (ii) $Score_{Max}(c) \stackrel{def}{=} \max_{d_i \in \mathcal{D}_{\text{init}}} p_{d_i}(q)$, (iii) $Score_{Min}(c) \stackrel{def}{=} \min_{d_i \in \mathcal{D}_{\text{init}}} p_{d_i}(q)$, (iv) $Score_{GeoMean}(c) \stackrel{def}{=} \sqrt[|c|]{\prod_{d_i \in c} p_{d_i}(q)}$, and (v) $Score_{HITS}(c)$. For the reference comparisons that incorporate free parameters — Rel Model, Rel Model(Re-Rank), and HITS— we employ leave-one-out cross validation so as to set free-parameter values as we do for ClustRanker. The performance numbers of all methods are presented in Table 10.

Our first observation based on Table 10 is that ClustRanker outperforms the initial ranking in most relevant comparisons; most of these improvements are quite substantial.

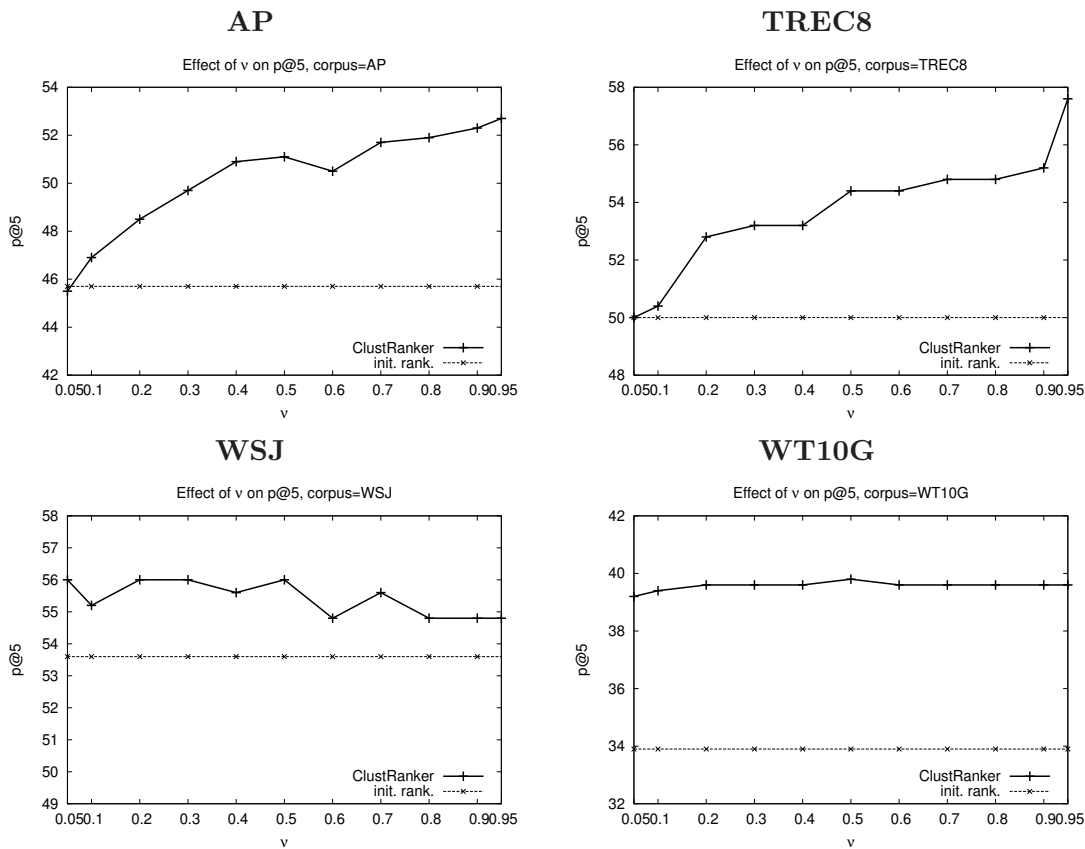

Figure 3: Effect of varying $\nu$, one of the graph parameters (refer to Appendix A), on the p@5 performance of ClustRanker.

The only exception is for the WSJ corpus, for which the relevance models also do not outperform the initial ranking in terms of p@5. Thus, it seems that query-variability issues, which affect the ability to learn effective free-parameter values, have quite an effect on the performance of methods for WSJ.

We can also see in Table 10 that except for WSJ, ClustRanker outperforms the previously proposed cluster ranking methods in almost all relevant comparisons. Many of these performance improvements are substantial and statistically significant.

As can be seen in Table 10, the ClustRanker method also outperforms each of the relevance models (Rel Model and Rel Model(Re-Rank)) in a majority of the relevant comparisons (corpus × evaluation measure). While there is a single case for which ClustRanker is outperformed in a statistically significantly manner by the relevance models (p@10 for WSJ), for WT10G ClustRanker posts statistically significant improvements over the relevance models, with some of the performance differences being quite striking. Furthermore, we observe that in contrast to ClustRanker, most previously proposed cluster ranking methods often post performance that is much worse — in many cases to a statistically significant degree — than that of the relevance models. Somewhat an exception is ranking clusters by the geometric mean of their constituent documents query-similarity values (GeoMean): for WT10G and WSJ the performance is better than that of the relevance models; however, for TREC8 and AP the performance is somewhat inferior to that of the relevance models.

|  | AP | | TREC8 | | WSJ | | WT10G | |
|---|---|---|---|---|---|---|---|---|
|  | p@5 | p@10 | p@5 | p@10 | p@5 | p@10 | p@5 | p@10 |
| init. rank. | 45.7 | 43.2 | 50.0 | 45.6 | 53.6 | 48.4 | 33.9 | 28.0 |
| Rel Model | 49.9 | **48.4**$^i$ | 50.8 | **50.2** | 52.0 | **52.6** | 30.4$^i$ | 29.2 |
| Rel Model(Re-Rank) | 51.1$^i$ | 45.8 | 50.4 | 49.8 | 52.0 | **52.6** | 36.3 | 27.8 |
| $S_{ClustQueryGen}(c) \stackrel{def}{=} p_c(q)$ | 39.2$^{ir\rho}$ | 38.8$^{ir\rho}$ | 39.6$^{ir\rho}$ | 40.6$^{ir\rho}$ | 44.0$^{ir\rho}$ | 37.0$^{ir\rho}$ | 30.0$^{\rho}$ | 24.1$^r$ |
| $S_{Max}(c) \stackrel{def}{=} \max_{d_i \in \mathcal{D}_{init}} p_{d_i}(q)$ | 41.8$^{r\rho}$ | 40.3$^{r\rho}$ | 38.8$^{ir\rho}$ | 41.6$^{r\rho}$ | 51.2$^{r\rho}$ | 46.6$^{r\rho}$ | 33.9 | 29.2 |
| $S_{Min}(c) \stackrel{def}{=} \min_{d_i \in \mathcal{D}_{init}} p_{d_i}(q)$ | 47.0 | 46.7 | 46.4 | 48.4$^{r\rho}$ | 48.4$^{r\rho}$ | 47.8$^{r\rho}$ | 31.4 | 25.9 |
| $S_{GeoMean}(c) \stackrel{def}{=} \sqrt[|c|]{\prod_{d_i \in c} p_{d_i}(q)}$ | 44.4$^{r\rho}$ | 46.7$^i$ | 50.0 | 49.6 | **56.0**$^{r\rho}$ | 50.6$^{r\rho}$ | 37.4$^r$ | 31.8$^{i\rho}$ |
| $S_{HITS}(c)$ | 48.5 | 47.2 | 50.8 | 43.2$^{r\rho}$ | 53.6 | 46.8$^{\rho}$ | 25.5$^{i\rho}$ | 23.9$^r$ |
| ClustRanker | **52.3**$^i_{cMg}$ | 48.3$^{ic}_M$ | **56.8**$_{cMmh}$ | 49.4$_{cM}$ | 53.2$_c$ | 46.2$^{\rho c}$ | **38.6**$^r_{cMm}$ | **31.2**$^{\rho}_{cm}$ |

Table 10: Performance results when using leave-one-out cross validation to set free-parameter values. Boldface marks the best performance per column. Statistically significant differences of a method with the initial ranking are marked with 'i'. Statistically significant differences of ClustRanker with the cluster ranking methods, ClustQueryGen, Max, Min, GeoMean, and HITS are marked with 'c', 'M', 'm', 'g', and 'h', respectively. Statistically significant differences of a cluster ranking method with Rel Model and Rel Model(Re-Rank) are marked with 'r' and '$\rho$', respectively.

All in all, the findings presented above attest that ClustRanker, when learning free parameter values, is (i) a highly effective method for obtaining high precision at top ranks, and (ii) much more effective than previously proposed methods for ranking clusters.

## 6. Conclusions and Future Work

We presented a novel language model approach to ranking *query-specific* clusters, that is, clusters created from documents highly ranked by some initial search performed in response to a query. The ranking of clusters is based on the presumed percentage of relevant documents that they contain.

Our cluster ranking model integrates information induced from the cluster as a whole unit with that induced from documents that are associated with the cluster. Two types of information are exploited by our approach: similarity to the query and centrality. The latter reflects similarity to other central items in the reference set, may they be documents in the initial list, or clusters of these documents.

Empirical evaluation showed that using our approach results in precision-at-top-ranks performance that is substantially better than that of the initial ranking upon which clustering is employed. Furthermore, the performance often transcends that of a state-of-the-art pseudo-feedback-based query expansion method, namely, the relevance model. In addition, we showed that our approach is substantially more effective in identifying clusters contain-ing a high percentage of relevant documents than previously proposed methods for ranking clusters.

For future work we intend to explore additional characteristics of document clusters that might attest to the percentage of relevant documents they contain; for example, cluster density as measured by inter-document-similarities within the cluster. Incorporating such characteristics in our framework in a principled way is an interesting challenge.

**Acknowledgments**

We thank the reviewers for their helpful comments. We also thank Lillian Lee for helpful comments on the work presented in this paper, and for discussions that led to ideas presented here; specifically, the cluster-centrality induction method is a fruit of joint work with Lillian Lee. This paper is based upon work supported in part by the Israel Science Foundation under grant no. 557/09, by the National Science Foundation under grant no. IIS-0329064, by Google's faculty research award, and by IBM's SUR award. Any opinions, findings and conclusions or recommendations expressed in this material are the authors' and do not necessarily reflect those of the sponsoring institutions.

## Appendix A. Centrality Induction

We briefly describe a previously proposed graph-based approach for inducing document centrality (Kurland & Lee, 2005), which we use for inducing document and cluster centrality.

Let $S$ (either $\mathcal{D}_{\mathrm{init}}$, the initial list of documents, or $Cl(\mathcal{D}_{\mathrm{init}})$, the set of their clusters) be a set of items, and $G = (S, S \times S)$ be the complete directed graph defined over $S$. The weight $wt(s_1 \to s_2)$ of the edge $s_1 \to s_2$ $(s_1, s_2 \in S)$ is defined as

$$wt(s_1 \to s_2) \stackrel{def}{=} \begin{cases} p_{s_2}(s_1) & \text{if } s_2 \in Nbhd(s_1; \delta), \\ 0 & \text{otherwise,} \end{cases}$$

where $Nbhd(s_1; \delta)$ is the set of $\delta$ items $s' \in S - \{s_1\}$ that yield the highest $p_{s'}(s_1)$. (Ties are broken by item ID.)

We use the PageRank approach (Brin & Page, 1998) to smooth the edge-weight function:

$$wt^{[\nu]}(s_1 \to s_2) = (1 - \nu) \cdot \frac{1}{|S|} + \nu \cdot \frac{wt(s_1 \to s_2)}{\sum_{s' \in S} wt(s_1 \to s')};$$

$\nu$ is a free parameter.

Thus, $G$ with the edge-weight function $wt^{[\nu]}$ constitutes an ergodic Markov chain, for which a stationary distribution exists. We set $Cent(s)$, the centrality value of $s$, to the stationary probability of "visiting" $s$.

Following previous work (Kurland & Lee, 2005), the values of $\delta$ and $\nu$ are chosen from $\{2, 4, 9, 19, 29, 39, 49\}$ and $\{0.05, 0.1, \ldots, 0.9, 0.95\}$, respectively, so as to optimize the p@k performance of a given algorithm for clusters of size $k$. We use the *same* parameter setting for the document-graph ($S = \mathcal{D}_{\mathrm{init}}$) and for the cluster-graph ($S = Cl(\mathcal{D}_{\mathrm{init}})$), and therefore inducing document and cluster centrality in any of our methods is based on two free parameters.

**Appendix B. Relevance Model**

To estimate the standard relevance model, RM1, which was shown to yield better performance than that of the RM2 relevance model (Lavrenko & Croft, 2003), we employ the implementation detailed by Lavrenko and Croft (2003). Let $w$ denote a term in the vocabulary, $\{q_i\}$ be the set of query terms, and $p_d^{JM[\alpha]}(\cdot)$ denote a Jelinek-Mercer smoothed document language model with smoothing parameter $\alpha$ (Zhai & Lafferty, 2001). RM1 is then defined by

$$p_{RM1}(w;\alpha) \stackrel{def}{=} \sum_{d \in \mathcal{D}_{\text{init}}} p_d^{JM[\alpha]}(w) \frac{\prod_i p_d^{JM[\alpha]}(q_i)}{\sum_{d_j \in \mathcal{D}_{\text{init}}} \prod_i p_{d_j}^{JM[\alpha]}(q_i)}.$$

In practice, RM1 is *clipped* by setting $p_{RM1}(w;\alpha)$ to 0 for all but the $\beta$ terms with the highest $p_{RM1}(w;\alpha)$ to begin with (Connell et al., 2004; Diaz & Metzler, 2006); further normalization is performed to yield a probability distribution, which we denote by $\tilde{p}_{RM1}(\cdot;\alpha,\beta)$. To improve performance, RM1 is anchored to the original query via interpolation using a free parameter $\gamma$ (Abdul-Jaleel et al., 2004; Diaz & Metzler, 2006). This results in the RM3 model:

$$p_{RM3}(w;\alpha,\beta,\gamma) \stackrel{def}{=} \gamma p_q^{MLE}(w) + (1-\gamma)\tilde{p}_{RM1}(w;\alpha,\beta);$$

$p_q^{MLE}(w)$ is the maximum likelihood estimate of term $w$ with respect to $q$. Documents in the corpus are then ranked by the minus cross entropy $-CE\left(p_{RM3}(\cdot;\alpha,\beta,\gamma) \,\|\, p_d^{Dir[\mu]}(\cdot)\right)$.

The free-parameter values are chosen from the following ranges to independently optimize p@5 and p@10 performance: $\alpha \in \{0, 0.1, 0.3, \ldots, 0.9\}$, $\beta \in \{25, 50, 75, 100, 500, 1000, 5000, ALL\}$, where "$ALL$" stands for using all terms in the corpus (i.e., no clipping), and $\gamma \in \{0, 0.1, 0.2, \ldots, 0.9\}$; $\mu$ is set to 2000, as in our cluster-based algorithms, following previous recommendations (Zhai & Lafferty, 2001).